\documentclass[RNAAS]{aastex63}

\begin{document}

\title{Uranus' stratospheric HCl upper limit from Herschel/SPIRE\footnote{{\it Herschel} is an ESA space observatory with science instruments provided by European-led Principal Investigator consortia and with important participation from NASA.}}

\correspondingauthor{N. A. Teanby}
\email{n.teanby@bristol.ac.uk}

\author[0000-0003-3108-5775]{N. A. Teanby}
\affiliation{School of Earth Sciences, University of Bristol, Wills Memorial Building, Queens Road, Bristol, BS8 1RJ, UK.}

\author[0000-0002-6772-384X]{P. G. J. Irwin}
\affiliation{Atmospheric, Oceanic \& Planetary Physics, Department of Physics, University of Oxford, Clarendon Laboratory, Parks Road, Oxford, OX1 3PU, UK.}

\begin{abstract}
Herschel/SPIRE observations of Uranus are used to search for stratospheric hydrogen chloride (HCl) emission at 41.74~cm$^{-1}$.
HCl was not detected and instead 3$\sigma$ upper limits were determined; $<$6.2~ppb ($<$2.0$\times$10$^{14}$~molecules/cm$^{2}$) for a 0.1~mbar step profile and $<$0.40~ppb ($<$1.2$\times$10$^{14}$~molecules/cm$^{2}$) for a 1~mbar step profile.
HCl is expected to have an external source and these upper limits are consistent with abundances of other external species (CO, H$_2$O, CO$_2$) and a solar composition source.
\end{abstract}


\section{Introduction} 

Hydrogen chloride (HCl) is a potentially important probe of external flux processes, interior composition, and atmospheric chemistry.
For giant planets, thermochemical models predict HCl is stable at high temperatures in the deep atmosphere \citep{94feglod}, but internally sourced HCl cannot reach observable tropospheric levels due to reactions with NH$_3$, which rapidly incorporate HCl into salts (NH$_4$Cl) at temperatures below $\sim$450--1500~K (pressures $<$200--10$^5$~bar) \citep{94feglod,01sho}.
Stratospheric conditions are more conducive to HCl survival, as at these altitudes NH$_3$ has been effectively removed by condensation, reactions with residual H$_2$S, and photolysis.
This implies any stratospheric HCl must be externally sourced \citep{01sho,21teaetal}.
External sources are inferred from detections of stratospheric water \citep{97feuetal}, CO \citep{14cavetal}, and CO$_2$ \citep{14ortetal}.
Sources include interplanetary dust particles (IDPs), volcanic moons, or large comet impacts.
Detection of HCl would provide an important independent probe of these external sources. 

Stringent HCl searches have been undertaken on Jupiter \citep{14teaetal_hcl}, Saturn \citep{12fleetal} and Neptune \citep{21teaetal}, but no detections have been made.
Lack of HCl may require loss mechanisms in addition to downward mixing, particularly for large cometary injections \citep{21teaetal}.
Possible loss mechanisms include aerosol scavenging \citep{01sho,14teaetal_hcl} or reactions with hydrocarbons \citep{21teaetal}.
It appears unlikely that HCl is detectable on Uranus, given its non-detection on Jupiter, Saturn, and Neptune.
However, Uranus is less hazy than Neptune \citep{20toletal} so aerosol scavenging may be less efficient.
HCl has a strong emission line at 41.74~cm$^{-1}$ that should be detectable if present in parts per billion (ppb) quantities.

\section{Observations}

Observations were taken with the Herschel space telescope \citep{10piletal} SPIRE spectrometer \citep{10grietal_full,10swietal}
and comprised a 8445~second Uranus integration starting 19:53UTC 10$^{\mbox{th}}$ July 2010 (Obs\_ID: 1342200175).
We used calibration v14.1.0 level 2 apodised radiances from SPIRE's short-wave spectrometer (SSW), covering 31.9--51.5~cm$^{-1}$ at 0.074~cm$^{-1}$ spectral resolution with a signal-to-noise of $\sim$1000 (Figure~\ref{fig:1}a).
Herschel's spatial resolution is $\sim$17$\arcsec$ at 41.74~cm$^{-1}$, which is large compared to Uranus' 3.57$\arcsec$ projected diameter, resulting in a disc-averaged spectrum. 

\section{Methods}

SPIRE observations were compared to disc-averaged synthetic spectra generated using the NEMESIS retrieval software \citep{08irwetal} following \citet{13teairw}.
Observations and synthetic spectra were converted into line-to-continuum ratios to allow direct comparison by selecting the 40--44~cm$^{-1}$ region and normalising with a low-order (quadratic) fit to the continuum after masking the HCl line position.
Using line-to-continuum ratios avoids issues with beam efficiency, fill factor, and baseline offsets.

\begin{figure}[ht]
\begin{center}
\includegraphics[scale=0.75,angle=0]{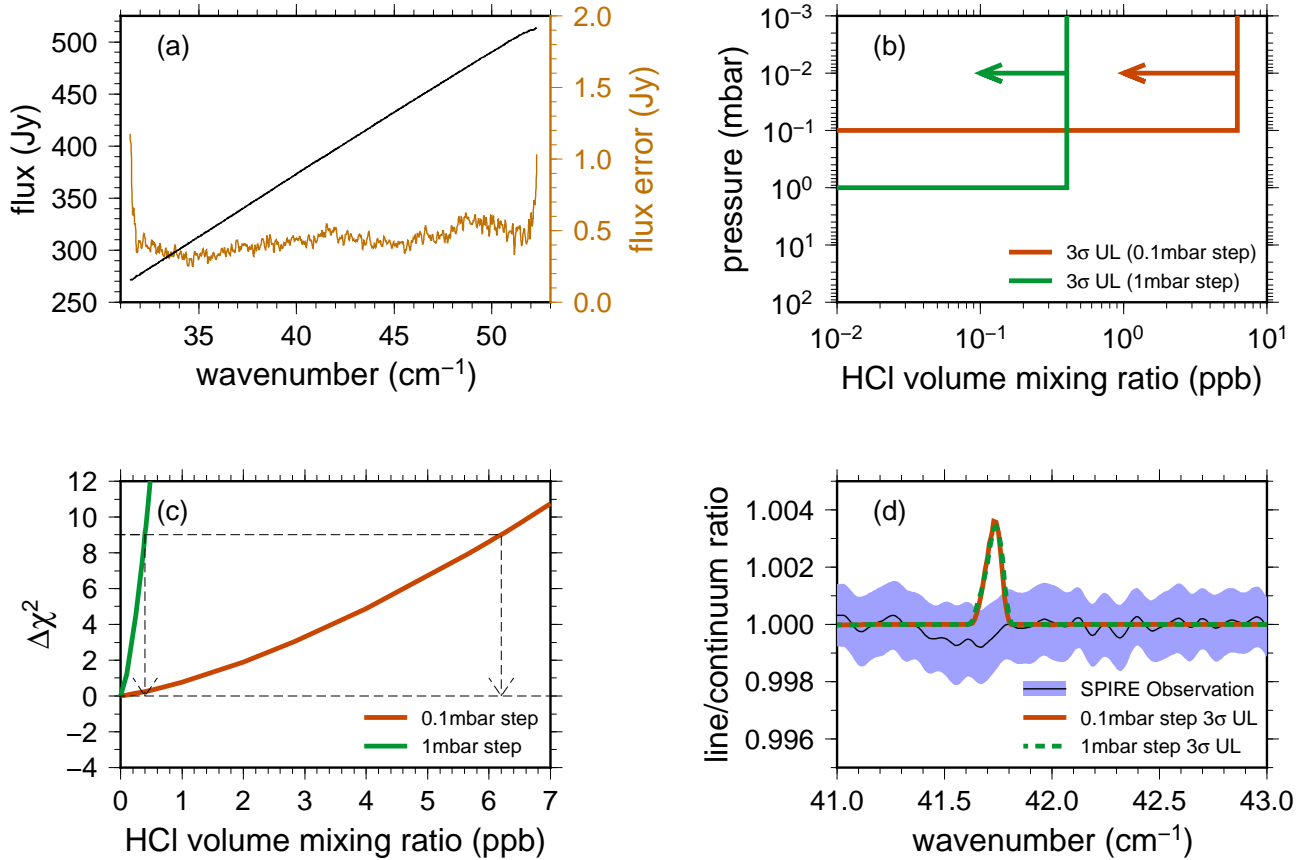}
\caption{(a) Herschel/SPIRE Uranus observation Obs\_ID:1342200175 and errors. (b) 3$\sigma$ HCl upper limit step-type profiles obtained using 0.1 and 1~mbar transition pressures. (c) $\Delta\chi^2$ misfit dependence on HCl VMR with 3$\sigma$ upper limits indicated. (d) Line-to-continuum ratio observation compared to synthetics generated using HCl upper limit abundances. \label{fig:1}}
\end{center}
\end{figure}

Following \citet{21teaetal} we used step-type HCl profiles, with uniform volume mixing ratios (VMRs) at pressures lower than a transition pressure and zero abundance at higher pressures (Figure~\ref{fig:1}b).
Transition pressures of 1 and 0.1~mbar were assumed, which are appropriate simple approximations for an external source \citep{21teaetal}. 
At higher pressures it is expected that aerosol scavenging and reactions with NH$_3$ removes HCl.
The $\chi^2$ misfit between observation and synthetics was determined as a function of HCl VMR $x$ following \citet{21teaetal}.
Detection at 3$\sigma$ significance requires $\Delta\chi^2=\chi^2(x)-\chi^2(0) \le -9$, or if no detection is made a 3$\sigma$ upper limit is defined by $\Delta\chi^2=+9$.

\section{Results}

Figure~\ref{fig:1}c shows $\Delta\chi^2(x)$ variation for 1 and 0.1~mbar step profiles.
No significant minima are present, so HCl is not detected on Uranus with these data.
Instead, SPIRE observations provide 3$\sigma$ upper limits of $<$6.2~ppb for 0.1~mbar step and $<$0.40~ppb for 1~mbar step, equivalent to column abundances of $<$2.0$\times$10$^{14}$ and $<$1.2$\times$10$^{14}$~molecules/cm$^{2}$ respectively.
Figure~\ref{fig:1}d compares the observation with synthetic spectra generated using these upper limits. 

\section{Discussion}

Our Uranus HCl upper limits can be compared to known externally sourced species (CO, H$_2$O, CO$_2$) to determine whether relative abundances are consistent with potential sources or if HCl non-detection requires other loss mechanisms.
Comparing column abundances is most robust as these are less sensitive to profile assumptions.
External oxygen species abundances are: 
H$_2$O=5--12$\times$10$^{13}$~molecules/cm$^{2}$ \citep{97feuetal};
CO$_2$=1.7$\pm$0.4$\times$10$^{13}$~molecules/cm$^{2}$ \citep{14ortetal}; and
CO=2.1--2.8$\times$10$^{17}$~molecules/cm$^{2}$ derived from \citet{14cavetal}'s 7.1--9.0~ppb 100~mbar step profile.
Therefore, CO is the major external oxygen carrier in Uranus' stratosphere and, when combined with our HCl upper limit, gives an upper limit on the external source's chlorine to oxygen ratio: Cl/O$<$4--9$\times$10$^{-4}$ (from HCl/CO).
This is consistent with Cl/O=3.3$\times$10$^{-4}$ for a solar composition source \citep{10lod}.
Additional loss processes such as aerosol scavenging \citep{01sho,14teaetal_hcl} or reactions with hydrocarbons \citep{21teaetal} may be occurring in Uranus' stratosphere, but they are not required to explain the HCl non-detection.

\section{Conclusion}

New Herschel/SPIRE 3$\sigma$ HCl upper limits on Uranus are $<$6.2~ppb ($<$2.0$\times$10$^{14}$~molecules/cm$^{2}$) for a 0.1~mbar step profile and $<$0.40~ppb ($<$1.2$\times$10$^{14}$~molecules/cm$^{2}$) for a 1~mbar step profile.
Therefore, HCl remains undetected on any of the four giant planets in our solar system.
A consistent interpretation of outer planet HCl upper limits and stratospheric CO abundances is that HCl is externally supplied by a combination of IDPs, comet impacts, or volcanic moons, but is removed from the stratosphere by aerosol scavenging or reactions with hydrocarbons.
For Jupiter and Neptune, these extra loss processes are required to explain non-detection of HCl if the external source is primarily large comets with solar composition.
Such comet impacts are required to explain large CO abundances and other shock chemistry products like CS in Jupiter and Neptune's stratosphere \citep{17moretal}.
For Saturn and Uranus, the upper limits are not stringent enough to require extra loss processes, but such processes are also likely on these planets.

\acknowledgments

NAT/PGJI are funded by the UK Science and Technology Facilities Council and UK Space Agency.
SPIRE has been developed by a consortium of institutes led by Cardiff University (UK) and including Univ. Lethbridge (Canada); NAOC (China); CEA, LAM (France); IFSI, Univ. Padua (Italy); IAC (Spain); Stockholm Observatory (Sweden); Imperial College London, RAL, UCL-MSSL, UKATC, Univ. Sussex (UK); and Caltech, JPL, NHSC, Univ. Colorado (USA). This development has been supported by national funding agencies: CSA (Canada); NAOC (China); CEA, CNES, CNRS (France); ASI (Italy); MCINN (Spain); SNSB (Sweden); STFC, UKSA (UK); and NASA (USA).

\end{document}